\documentstyle[aas2pp4]{article}
\input epsf.tex

\lefthead{Novosyadlyj \& Chornij}
\righthead{Spatial Correlation Function of Quasars}

\begin{document}
\def\solar{\ifmmode_{\mathord\odot}\else$_{\mathord\odot}$\fi}

\begin{flushleft}
\small KiFNT, v. 14, No 2(1998) p.156-165
\end{flushleft}

\title{\bf SPATIAL CORRELATION FUNCTION OF QUASARS AND
 POWER SPECTRUM OF COSMOLOGICAL MATTER DENSITY PERTURBATIONS}
\author{B. Novosyadlyj, Yu. Chornij}
\affil{ Astronomical Observatory of L'viv State University,
 Kyrylo i Methodij str.8, 290005 L'viv, Ukraine}



\begin{abstract}
We examine the dependence of the spatial two-point correlation
function of quasars $\xi_{qq}(r,z)$ at different redshifts
on the initial power spectrum in flat cosmological models. Quasars
and other elements of the large-scale structure of the universe are
supposed to form in the peaks of the scalar Gaussian field of density
fluctuations of appropriate scales. Quasars are considered as a manifestation
of short-term active processes at the centers of these fluctuations;
such processes set in when dark matter counterflows and a shock wave
appear in the gas. We propose a method for calculating the correlation
function $\xi_{qq}(r,z)$ and show its amplitude and slope
to depend on the shape of the initial power spectrum and the scale
$R$ of the fluctuations in which quasars are formed. We demonstrate
that in the CDM models with the initial power spectrum slope 
\mbox{$n=0.7\div1$}
it is possible to explain, by choosing appropriate
values of $R$, how the amplitudes and correlation radii of 
$\xi_{qq}(r,z)$ may either increase or decrease with increasing
redshift $z$. In particular, the correlation radii of 
$\xi_{qq}(r,z)$ grow from \mbox{$6-10h^{-1}$ Mpc} when $R$ grows from 
\mbox{$0.45 h^{-1}$} to \mbox{$1.3h^{-1}$ Mpc}. The H+CDM 
model at realistic values of $R$ fails to account for the observational
data according to which the $\xi_{qq}(r,z)$ amplitude decreases
with increasing $z$.
\end{abstract}


\section{Introduction}
 The most elaborated scenario of the origin of the large-scale
structure in the universe is pictured by the model in which such a
structure results from the evolution of the uniform isotropic Gaussian
scalar field of cosmological matter density fluctuations under the
effect of gravitational instability. The major unresolved problems
in this case are the choice of an inflationary model for an early
universe, determination of the nature of the "dark" non-baryon
matter (DM) and its fraction in the total matter mass. Assumptions
as to the inflationary model and DM nature define the shape of the
initial (post-recombination) power spectrum of cosmological density
fluctuations, $P(k)$, and thus the principal parameters of large-scale
structure can be theoretically calculated. Therefore it is of importance
to test cosmological models with given power spectra $P(k)$. This
testing can be done by calculating spatial two-point correlation functions
for large-scale structure elements on various scales and comparing
them with observational data. The testing is based on the relationship
between the characteristics of structure elements and their correlation
functions, on the one hand, and the amplitude and slope of the initial
spectrum $P(k)$ at different $k$, on the other hand. Information
on the power spectrum on small and intermediate scales 
(\mbox{$r\le 100h{-1}$ Mpc}, \mbox{$h = H_0/(100 km\cdot s^{-1}Mpc^{-1}$},
$H_0$ being the Hubble constant at the present epoch) resides
in the correlation functions of bright massive galaxies, rich clusters
of galaxies, and quasars. The "observed" correlation functions
of all these three types of objects are described by the approximate
expression \mbox{$\xi(r) = (r/r_0)^{-1.8}$}, where $r_0$
is the correlation radius equal to 
\mbox{$5.4h^{-1}$ Mpc}, \mbox{$18h^{-1}$ Mpc} [\cite{bs}-\cite{iras}],
\mbox{$(6\div10)h^{-1}$ Mpc} [\cite{adr}-\cite{mo}]. 
for the three types of objects, respectively. At the moment investigations
of the "observed" correlation function of quasars in different
redshift ranges give ambiguous results. For example, the correlation
function amplitude found in [\cite{iov2,komb2,mo}] decreases with increasing $z$,
it remains unchanged in [\cite{adr}], and grows at \mbox{$z\le 1.7$} 
and diminishes at \mbox{$z>1.7$} in [\cite{komb3}]. The common result
of these studies is that the amplitude for quasars is larger than
for galaxies but is smaller than for rich clusters of galaxies. This
result was obtained by different authors from two quasar samples:
a combined sample of all quasars observed in the $z$ range from
$0.1$ to $4.5$ and the "nearby" quasars at \mbox{$z\le 1.5$}.

Here we look into the possibility of using the above results as a
test for cosmological models with given power spectra $P(k)$; to
this end, the theoretical correlation function of quasars has to be
calculated.

Theoretical methods for calculating correlation functions of galaxies
and their clusters are based on the theory of Gaussian random fields
[\cite{dor2}-\cite{kais}], they have been devised in detail. 
The results obtained within
the scope of cosmological models with given initial power spectra
were analyzed in [\cite{hn1}-\cite{watan}], for example. 
Calculation of the correlation
function of quasars $\xi_{qq}$ is complicated by a number
of problems which have yet to be resolved. What scale is typical of
the regions where quasars formed at different $z$? What are typical
physical parameters of quasars: mass, duration of formation, lifetime,
etc.? What is the relation between $\xi_{qq}$ and these
parameters? Answers to these questions essentially depend on the physical
model chosen for the quasar phenomenon. In particular, the disk accretion
of gas on a massive black hole at the center of a galaxy may be a
model mechanism. Therefore, for the mass of the fluctuations in which
quasars are formed we can take the mass of "parent" galaxy
$M_{g/q}$ which is able to ensure a high luminosity
of the nucleus (observed as a quasar) during the quasar lifetime 
$\tau_q$. Based on the results of [\cite{efst}-\cite{turn}], we may take
\mbox{$10^{11}-10^{12}M\solar$}  for $M_{g/q}$
(the black hole mass \mbox{$\sim10^8-10^9M\solar$}),
\mbox{$\sim 10^7-10^8$} yr for $\tau_q$, 
and 
\mbox{$\Delta t_q\le10^8$} yr for
the duration of quasar formation. 
Whether these parameters are the same
for quasars at different $z$ is still an open question.

\section{Principal assumptions and formulation of the problem}
In the cosmological scenario used by us, galaxies, rich clusters
of galaxies, and quasars appear in the peaks of the scalar Gaussian
field of matter density fluctuations on corresponding scales, the
relative amplitude of fluctuations being 
\mbox{$\delta\equiv\delta\rho/\rho=\sigma\cdot\nu$}
($\sigma$ is the rms amplitude, $\nu$ is the peak
height). It is assumed that galaxies and their clusters come into
being when counterflows arise in the DM and a shock wave arises in
the gas. The amplitude $\delta$ at this moment $t$ is determined
from Tolmen's model in terms of redshift: 
\mbox{$\delta(z) =1.69 \cdot (z+1)$}.
The amplitude corresponding to the objects that appeared earlier
is \mbox{$\delta>\delta (z)$},
it has a normal distribution: 
\mbox{$p(\delta)=(2\pi\sigma^2)^{-1/2}\cdot exp(-\delta^2/2\sigma^2)$}.
The probability that a galaxy or
a rich cluster of galaxies occurs at a fixed $z$ is
\begin{equation}
\label{P1}
P_1(z)=\int\limits_{\delta(z)}^{\infty} p(\delta)\,d\delta.
\end{equation}
The probability that two galaxies exist simultaneously at
a fixed $z$ at two different points 
$\vec x_1$ and $\vec x_2$ ($r=|\vec x_1-\vec x_2|$)
is
\begin{equation}
\label{P2}
P_2(z)=\int\limits_{\delta(z)}^{\infty}\int\limits_{\delta(z)}^{\infty} p(\delta_1,\delta_2)\,d\delta_1d\delta_2,
\end{equation}
where 
$p(\delta_1,\delta_2)$ is the two-dimensional
normal distribution of random amplitudes 
$\delta_1$ and $\delta_2$ [\cite{ven}]:
$$p(\delta_1,\delta_2)=(2\cdot\pi)^{-1}\cdot\left(\sqrt{\xi^2(0,z)-\xi^2(r,z)}\right)^{-1}$$
\begin{equation}
\label{p2}
\times exp\left(-\frac{\xi(0,z)\cdot\delta_1^2+\xi(0,z)\cdot\delta_2^2-2\cdot\xi(r,z)\cdot\delta_1\cdot\delta_2}{2\cdot\left(\xi^2(0,z)-\xi^2(r,z)\right)}\right),
\end{equation}
here 
$r>0$ and $\xi(r,z)$ is the correlation function
of the density fluctuations in which the objects are formed. The function
is calculated from the given initial power spectrum 
$P\left(k,R_f\right)$ 
smoothed on the scale 
$R_f$ 
which corresponds to the scale of the objects:
\begin{equation}
\label{xi_p}
\xi(r,z)=\frac{1}{2\pi^2}\cdot\int\limits_0^{\infty}\\
k^2\cdot\frac{P\left(k,R_f\right)}{(1+z)^2}\cdot\frac{sin(kr)}{kr}\,dk,
\end{equation}
where 
$$P\left(k,R_f\right)=P(k)\cdot W^2(kR_f)$$ and 
$$W(kR_f)=exp\left(-\frac{1}{2}\cdot k^2\cdot R_f^2\right)$$
is the smoothing function.
The statistical correlation function
of the fluctuation peaks in which cosmological objects are formed
is, by definition,
\begin{equation}
\label{xi_oo_o}
\xi_{oo}^{st}(r,z)\equiv\frac{P_2(z)}{P_1^2(z)}-1.
\end{equation}
This function for rich clusters of galaxies or for galaxies
at $z=0$ is [\cite{kais}]
$$\xi_{oo}^{st}(r)\equiv\xi_{oo}(r,z=0)=\sqrt{\frac{2}{\pi}}\cdot\left(erfc\left(\frac{\nu}{\sqrt{2}}\right)\right)^{-2}\times$$
\begin{equation}
\label{xi_oo}
\times\int\limits_{\nu}^{\infty}\\
e^{-1/2\cdot y^2}\cdot erfc\left(\frac{\nu- y\cdot \xi(r)/\xi(0)}{\sqrt{2\cdot(1-\xi^2(r)/\xi^2(0))}}\right)\,dy-1.
\end{equation}

Expression (6) is simplified at 
\mbox{$\xi(r)\ll1$}  and \mbox{$\nu\gg1$}
\begin{equation}
\label{xi1_oo}
\xi_{oo}^{st}(r)\approx \left(\frac{\nu}{\sigma}\right)^2\cdot \xi(r).
\end{equation}

In this case $\xi_{oo}^{st}(r)$ for objects
is related to $\xi(r)$ for the fluctuations on the corresponding
scale in which the objects are formed. The factor \mbox{$\nu/\sigma$}
is called the statistical biasing of these objects, and the procedures
for its determination for galaxies and rich clusters of galaxies are
described in [\cite{bbks}-\cite{hn1,co_ka}]. In deriving the correlation function of
quasars we take advantage of the approach proposed in [\cite{kais}] with allowance
made for the specific nature of quasars.

\section{Correlation function of quasars}

To calculate correlation functions of galaxies and rich clusters
of galaxies, we have to specify the scale of the corresponding fluctuations
in which they are formed. For quasars, there are two more important
quantities along with this scale: the duration of quasar formation
$\Delta t_{q}$ and quasar lifetime $\tau_{q}$.
The smallest mass of the galaxies which may experience the quasar
phase in the course of their evolution is 
\mbox{$M_{g/q}^{min}\sim 2\cdot10^{11}M\solar$} [\cite{efst}-\cite{turn}].
The duration \mbox{$\Delta t_{q}\le10^8$} yr is
much shorter than the cosmological evolution time of fluctuations
in which such galaxies are formed at \mbox{$z\le5$}, and so
we may ignore this quantity in first approximation. The same studies
[\cite{efst}-\cite{turn}] reveal that the quasar phase is short-lived 
\mbox{$\tau_{q}\sim 10^7-10^8$} yr. The quantities
$P_1(z)$ and $P_2(z)$ in (1), (2) are the probabilities
that the random amplitudes of corresponding fluctuations are 
\mbox{$\delta>\delta(z)$}, i. e., they determine the probability of coming
into being or the probability of the existence of objects on
corresponding scales with redshifts larger than  a given $z$.  For
quasars, the interval $\Delta(z,\tau_q)$ between the limits of integration
in (1) and (2) is determined as the lifetime $\tau_q$
of the quasars which came into being at the cosmological moment 
$t(z)$ corresponding to a given $z$ [\cite{nov3}-\cite{nov_qso}]:
\begin{equation}
\label{P2_2}
\Delta(z,\tau_q)=1.13\cdot (z+1)\cdot\frac{\tau_q}{t(z)}.
\end{equation}
When $\tau_q\sim t(z)$, the quantities
$P_1(z)$ and $P_2(z)$ are in form (1), (2) with the
upper limit of integration $\delta(z)+\Delta(z,\tau_q)$.
When $\tau_q\ll t(z)$, expressions (1), (2) take the form
\begin{equation}
\label{P1_1}
P_1^q(z)=p(\delta(z))\cdot \Delta(z,\tau_q),\; P_2^q(z)=p(\delta_1,\delta_2)\cdot\Delta^2(z,\tau_q).
\end{equation}
in view of smallness of $\Delta(z,\tau_q)$. Then, on the basis of
(5) and in view of (9), the correlation function of quasars is
$$\xi_{qq}^{st}(r,z)=\left(\sqrt{1-\left(\frac{\xi(r)}{\xi(0)}\right)^2}\right)^{-1}$$
\begin{equation}
\label{xi_ªª}
\times exp\left(\frac{2.86\cdot(z+1)^2}{\xi(0)}\cdot\left(1+\frac{\xi(0)}{\xi(r)}\right)^{-1}\right)-1,
\end{equation}
As it follows from (4), the ratio \mbox{$\xi(r,z)/\xi(0,z)$}
is independent of $z$, and \mbox{$\xi(r)\equiv\xi(r,z=0)$}.

The above expression for $\xi_{qq}^{st}(r,z)$ represents
the statistical component in the correlation function of the fluctuation
peaks where quasars are formed, but it takes no account of the dynamics
of background large-scale inhomogeneities (the dynamical component).
According to [\cite{bbks}], the complete correlation function of objects is
\begin{equation}
\label{xi_k_p}
\xi_{oo}(r,z)=\left(\sqrt{\xi_{oo}^{st}(r,z)}+\sqrt{\xi_{oo}^{d}(r,z)}\right)^2,
\end{equation}
where the statistical component for quasars 
$\xi_{oo}^{st}(r,z)\equiv \xi_{qq}^{st}(r,z)$
is represented by (10) and the dynamical component 
\mbox{$\xi_{oo}^{d}(r,z)\equiv\xi(r,z)$}  is correlation function
(4) of density fluctuations.

Let us examine the approximation 
(\mbox{$\xi(r)/\xi(0))\ll1$}. In this approximation $\xi_{qq}^{st}(r,z)$
(10) takes the form
\begin{equation}
\label{xi_ªª_t}
\xi_{qq}^{st}(r,z)\approx\left(\frac{1.69\cdot(1+z)}{\sigma^2}\right)^2\cdot\xi(r)\\
=\left(\frac{\nu}{\sigma}\right)^2\cdot\xi(r),
\end{equation}
after its expansion into the Taylor series. It coincides
with expression (7) for the correlation function of the galaxies and
clusters which formed in high peaks \mbox{$\nu\gg1$}
(that is, at high redshifts), but in the case of quasars there is no such a
restriction to high peaks only.

According to (4), the dynamical component $\xi(r,z)$ is 
expressed in terms of $\xi(r)$
\begin{equation}
\label{xi_d}
\xi(r,z)=\xi(r)\cdot(1+z)^{-2}.
\end{equation}
Then, in view of (11)-(13), we can write
\begin{equation}
\label{xi_kk_tp}
\xi_{qq}(r,z)=\left(\frac{1.69\cdot(1+z)}{\sigma^2}+\frac{1}{1+z}\right)^2\cdot\xi(r).
\end{equation}
As seen from (12) and (13), the statistical component in
the correlation function of quasars increases with $z$, while
the dynamical component diminishes. Therefore, when $z$ is variable
and $r$ is fixed, the correlation function amplitude may have
a minimum if a decrease in the dynamical component is not compensated
by an increase in the statistical component at small $z$. The
minimum point can be found from the condition 
\mbox{$\partial \xi_{qq}(r,z)/\partial z=0$}
with expression (14):
\begin{equation}
\label{extr}
z_*=\frac{\sigma}{1.3}-1.
\end{equation}
Thus, at $\sigma<1.3$ we have $\partial \xi_{qq}(r,z)/\partial z>0$
within the interval $z>0$ the correlation
function amplitude does nothing but grows. 
When \mbox{$\sigma>1.3$},
the amplitude is minimum at \mbox{$z=z_*$}. Over
the entire range \mbox{$z\le5$}, in which quasars are observed,
the amplitude of the correlation function of quasars diminishes at
\mbox{$\sigma\geq7.8$}, over the range \mbox{$z\le4$}
it diminishes at \mbox{$\sigma\geq6.5$}, and over the range \mbox{$z\le3$}
at \mbox{$\sigma\geq5.2$}. These results were derived in the
approximation \mbox{($\xi(r)/\xi(0))\ll1$}, which is
valid at \mbox{$r\geq10h^{-1}$ Mpc}. On these scales,
the large-scale inhomogeneities have the amplitude 
\mbox{$<(\delta\rho)/\rho>\ll1$} even at
\mbox{$z\sim0$}, so that the correction of the spectrum
$P(k)$ for the nonlinear evolution does not affect the results.

\section{Calculation results}

We calculated the correlation function of quasars $\xi_{qq}(r,z)$
within the framework of models with various initial
power spectra of fluctuations $P(k)$. Among models with various
$\Omega_{b}$
(fraction of baryon density in the density of the whole matter
in terms of the critical density \mbox{$\Omega_b=\rho_{b}/\rho_{cr}$)}, 
$\Omega_{CDM}$, and $\Omega_{HDM}$
(CDM stands for the cold dark matter of axion
type and HDM for the hot dark matter of massive neutrino type), the
CDM models with \mbox{$\Omega_{b}\le0.1$} and 
$\Omega_{CDM}=1-\Omega_{b}$  are believed
to be the most promising at the moment; they include the "standard"
CDM model with the spectrum slope \mbox{$n=1$} 
($P(k)=A\cdot k^n\cdot T^2(k)$,
where $A$ is a spectrum normalization constant,
$k$ is the wave number, and $T(k)$ is a transfer function
depending on DM nature), "inclined" CDM models with 
\mbox{$n=0.8;0.7$}, and the "hybrid" H+CDM model with 
\mbox{$\Omega_{b}=0.1$}, 
\mbox{$\Omega_{CDM}=0.6$}, \mbox{$\Omega_{HDM}=0.3$}, \mbox{$n=1$}.
Model parameters and methods of spectrum
normalization on the COBE results [\cite{benn,smoot}] were described 
in [\cite{nov2,nov_qso,nov_tr}].
To calculate $\xi_{qq}(r,z)$, one has to specify the characteristic
scale or mass of fluctuations in which quasars are formed and smooth
the power spectrum $P(k)$ by a Gaussian filter with the radius
$R_f$ corresponding to this scale. It follows from the expression
\mbox{$M=4.37\cdot10^{12}R_f^3h^{-1}M\solar$} [\cite{bbks}] 
that the radius corresponding to the mass 
\mbox{$M_{g/q}^{min}\simeq 2\cdot10^{11}M\solar$} 
[\cite{efst}-\cite{turn}] is \mbox{$R_f\simeq 0.35h^{-1}$ Mpc}. 
So, we start from the assumption 
that quasars are an early short phase in the evolution of galaxies
with this mass and calculate the correlation function of quasars at
this value of $R_f$.

Figure 1 shows approximations of the observed correlation functions
of galaxies [\cite{dav,iras}] 
and rich clusters of galaxies [\cite{bs,iras}] and the correlation
function of quasars calculated by expression (11) for 
$z=0.5, 1, 2, 3, 4$ in various cosmological models.

\begin{figure}[p]
\epsfxsize=7truecm
\centerline{\epsfbox{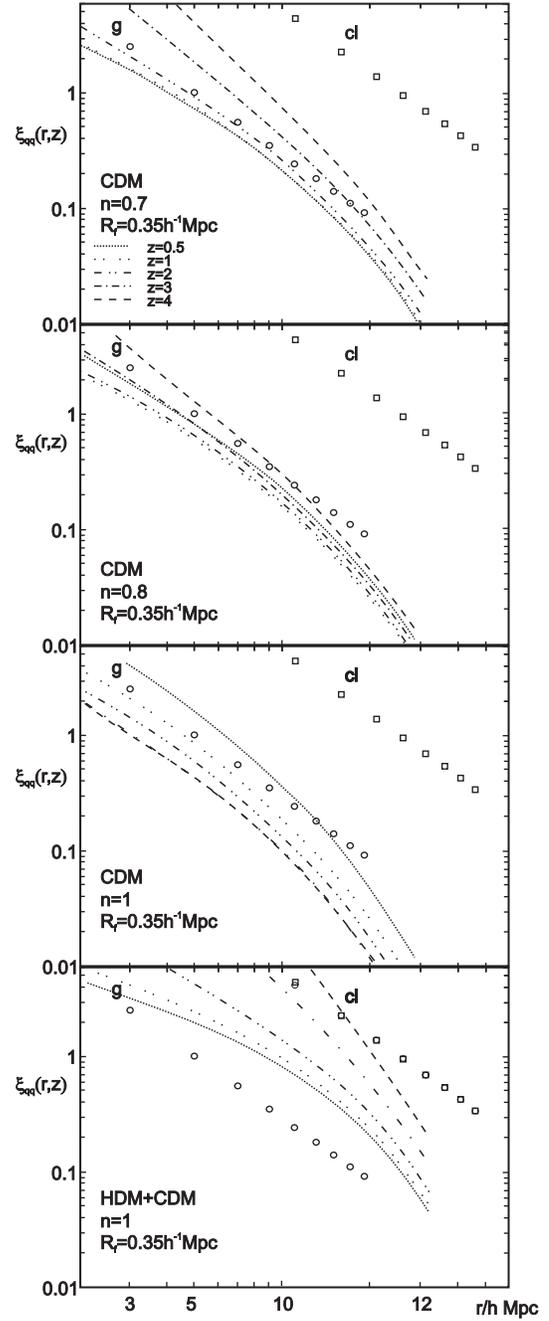}}
\caption{
Correlation functions of quasars $\xi_{qq}(r,z)$ 
calculated in different models at $R_f=0.35h^{-1}$ Mpc 
and correlation functions of galaxies (circles) and rich clusters 
of galaxies (squares) approximated by the expression 
$\xi(r)=(r/r_0 )^{-1.8}$ at $r_0^g \approx 5.4 
h^{-1}$ Mpc  and $r_0^{cl}\approx 18h^{-1}$ Mpc.
}
\end{figure}
It is apparent that at large redshifts, where the correlation function
is mainly specified by its statistical component and the dynamical
component is of minor importance, the amplitudes of correlation functions
of quasars diminish with increasing $n$ in various CDM models.
This is a manifestation of the well-known tendency of Gaussian field
peaks to decrease the degree of their clustering with decreasing height
$\nu$. Calculations of these heights for a fixed $z$ by the
expression 
\mbox{$\nu=\delta(z)/\sigma=1.69\cdot(1+z)/\sigma$}
where \mbox{$\sigma\equiv\sigma(R_f)=\sqrt{\xi(0)}$} (4), 
reveal that they are the largest in the CDM model with
$n = 0.7$ ($\sigma=1.67$), somewhat smaller in the CDM model
with $n=0.8$ ($\sigma=2.37$), and the smallest in the CDM
model with $n=1$ ($\sigma=4.78$). The above rms amplitudes
$\sigma$ allow one to follow the variations of correlation function
amplitudes for quasars in these models at different $z$ (see
expression (15)). Thus, the amplitude $\xi_{qq}(r,z)$ in the CDM model with 
$n=0.7$ diminishes at small redshifts, reaches its minimum at $z$
$z\approx0.3$, and then it grows. In the CDM model with $n=0.8$,
a similar minimum occurs at $z\approx0.8$, in
the CDM model $n = 1$ at $z\approx2.7$, and in the
H+CDM ($\sigma=1.39$) - at $z\approx0.07$.
These results
are in good agreement with numerical calculations (Fig. 1). When the
plots in Fig. 1 are matched to observational data 
[\cite{adr}-\cite{mo}],
one can see that only the CDM model with $n = 1$ and results
of [\cite{iov2}] are in accord as to the correlation radius 
\mbox{$r_0(z=0.5)\sim(6\div10)h^{-1}$ Mpc} and
the trend of the $\xi_{qq}(r,z)$ amplitude variations.
As regards other observational data, the cosmological models discussed
here under the assumption that quasars are formed in the fluctuations
on the scale \mbox{$R_f=0.35h^{-1}$ Mpc} fail to explain the
amplitudes of the observed correlation functions of quasars at various
$z$ and their correlation radii. According to 
[\cite{adr}-\cite{mo}], the
radii \mbox{$r_0(z)\sim(6\div10)h^{-1}$ Mpc}
at different redshifts. These correlation function parameters in the
CDM models are smaller than in the observed function, while in the
H+CDM model they are too large. This may mean that the fluctuation
scale chosen by us, \mbox{$R_f=0.35h^{-1}$ Mpc}, is too small
in the CDM models and too large in the H+CDM model.
The correlation functions $\xi_{qq}(r,z)$ plotted in Fig. 2 were
calculated with the CDM models with such fluctuation scales 
$R_f$ for each model that the functions do not contradict the data of
[\cite{adr,komb2,mo}] on the amplitude (or correlation radius) on the scales 
\mbox{$r\leq 12h^{-1}$ Mpc}. At \mbox{$12< r\leq 40\,h^{-1}$ Mpc} 
the amplitudes of these correlation
functions of quasars are no smaller than the amplitude of the observed
correlation function of galaxies, with due regard for a large scatter
of that function on these scales. 
\begin{figure}[p]
\epsfxsize=7truecm
\centerline{\epsfbox{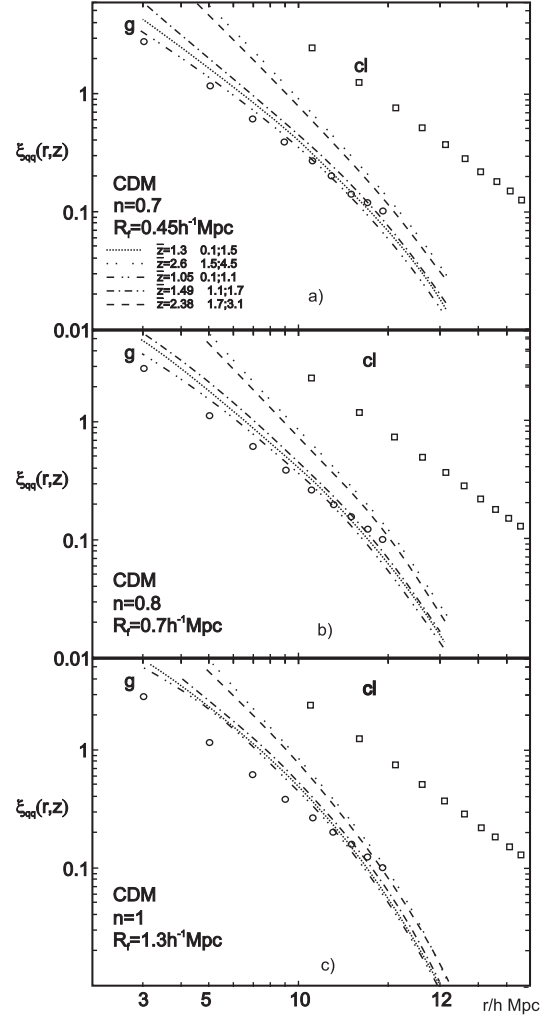}}
\caption{
Correlation functions of quasars $\xi_{qq}(r,z)$ 
calculated in CDM models for $\bar z=1.3, 2.6, 1.05, 1.49, 2.38$ 
weighted means in the $\Delta z$ intervals 
$0.1-1.5$, $1.5-4.5$, $0.1-1.1$, $1.1-1.7$, $1.7-3.1$. The 
scale $R_f$ is chosen such that the correlation radii may be 
within the range $6-10h^{-1}$ Mpc. 
Other designations are the same as in Fig. 1. 
}
\end{figure}
Besides, to make more correct the
comparison between the theoretical and observational data, we took
into consideration that the observed correlation functions of quasars
were determined for wide redshift intervals 
\mbox{$\Delta z\equiv [z_{min},z_{max}]$}
  [\cite{adr}-\cite{komb2,mo}], while
the theoretical functions were calculated for fixed $z$. Therefore,
the observed correlation functions at different $\Delta z$ were fitted
by the theoretical functions calculated for $\bar z$, the weighted
mean in these intervals found from the observed distribution of quasars
$n_q(z)$ [\cite{schmidt}] in the comoving reference frame:
$$\bar z= \frac{\int\limits^{z_{max}}_{z_{min}}z\cdot n(z)\cdot r^2(z)\,dr(z)}
{\int\limits^{z_{max}}_{z_{min}} n(z)\cdot r^2(z)\,dr(z)},$$
where $r(z)=2c/H_0\cdot\left(1-(z+1)^{-0.5}\right)$
is the distance in the comoving reference frame, 
$dr(z)=c/H_0\cdot(z+1)^{-1.5}\,dz$, 
and $c$ is the 
velocity of light. The weighting made to fit $\xi_{qq}(r,z)$
to the results of [\cite{iov2,mo}] gave 
$\bar z=1.3$ for the interval $\Delta z\equiv[0.1,1.5]$ 
and $\bar z=2.6$ for $[1.5,4.5]$.
For the intervals $\Delta z$ used in studies [\cite{adr,komb2}] we found 
$\bar z=1.05$ for $[0.1,1.1]$, $\bar z=1.49$ for $[1.1,1.7]$, and
$\bar z=2.38$ for $[1.7,3.1]$.

Based on the correlation functions of quasars which correspond to
the data of [\cite{adr,komb2,mo}] as to correlation radius 
\mbox{$r_0^q(z)\sim 6 \div 10h^{-1}$ Mpc},
we determined the scales $R_f$ of fluctuations
in which quasars are formed and their rms amplitudes $\sigma(R_f)$
in the CDM models:
$R_f\approx1.3h^{-1}$ Mpc, $\sigma(R_f)=2.24$ 
(in the model with $n=1$); 
$R_f\approx0.7h^{-1}$ Mpc, $\sigma(R_f)=1.73$ ($n = 0.8$); 
$R_f\approx0.45h^{-1}$ Mpc, $\sigma(R_f)=1.52$ ($n = 0.7$). 

Judging from the above values of $\sigma(R_f)$
and expression (15), the amplitude and the correlation radius of 
$\xi_{qq}(r,z)$ in these models grow with $\bar z$.

\section{Conclusion}
We have elaborated a method for the theoretical calculation
of the correlation function of quasars, it is based on the Gaussian
statistics of the initial field of cosmological matter density fluctuations.
The correlation functions of quasars were calculated for different
redshifts in cosmological models with given initial power spectra
$P(k)$ on the assumption that quasars are formed in the peaks of
the fluctuations and exist over a time much shorter than the cosmological
time. The amplitudes and slopes of the correlation functions are shown
to depend on fluctuation scale and power $P(k)$ on corresponding
scales. The amplitudes and correlation radii of the observed and theoretical
correlation functions of quasars are consistent in the CDM models
with the spectrum slope $n$ ranging from 0.7 to 1 when the scale
$R_f$ of the corresponding fluctuations in which quasars are
formed ranges from \mbox{$0.45h^{-1}$ Mpc} to \mbox{$1.3h^{-1}$ Mpc}.
In this case, however, the function amplitudes increase with $z$.
Although there is no evidence of any monotonic growth of the correlation
function amplitude in studies [\cite{adr,iov2,komb2,komb3,mo}], 
the estimates are so
contradictory that we cannot use them to test the cosmological models.
The H+CDM model fails to explain the correlation radius derived from
observational data \mbox{($6h^{-1}\div 10h^{-1}$ Mpc)}
if the scale of fluctuations in which quasars are formed exceeds 
\mbox{$0.35h^{-1}$ Mpc}. Our results are in accord with the physical models
which regard the quasar phenomenon as an early short phase in the
evolution of massive galaxies or as a merger of galaxies situated
in groups.


\end{document}